\documentclass[twocolumn,english,aps,prl,preprint,reprint,showpacs,longbibliography,showkeys]{revtex4-1}
\usepackage[T1]{fontenc}
\usepackage{lmodern}
\usepackage{color}
\usepackage{babel}
\usepackage{amsmath}
\usepackage{amssymb}
\usepackage{graphicx}

\RequirePackage[colorlinks=true,linkcolor=blue]{hyperref}
\usepackage{dsfont}
\usepackage{cleveref}
\usepackage{natbib}
\usepackage{bm}

\makeatletter
\usepackage{amsfonts}
\usepackage{babel}
\usepackage{braket}
\usepackage{verbatim}
\usepackage{amstext}
\makeatother

\begin{document}

\title{Input-output theory for spin-photon  coupling in Si double quantum dots}

\author{M. Benito,$^{1}$ X. Mi,$^{2}$ J. M. Taylor,$^{3}$ J. R. Petta,$^{2}$  and Guido Burkard$^{1}$}

\address{$^{1}$Department of Physics, University of Konstanz, D-78457 Konstanz, Germany}

\address{$^{2}$Department of Physics, Princeton University, Princeton, New Jersey 08544, USA}

\address{$^{3}$Joint Quantum Institute/NIST, 
College Park, Maryland 20742, USA}

\begin{abstract}
The interaction of qubits via microwave frequency photons 
 enables long-distance qubit-qubit coupling
 and facilitates the realization of a large-scale quantum processor.
However, qubits based on electron spins in semiconductor quantum dots
 have proven challenging to couple to microwave photons.
In this theoretical work we show that a sizable coupling for a single electron spin is possible via spin-charge hybridization using a magnetic field gradient  in a silicon double quantum dot.
Based on parameters already shown in recent experiments, we predict optimal working points to achieve a coherent spin-photon coupling, an essential ingredient for the generation of long-range entanglement.
Furthermore, we employ input-output theory to identify  observable signatures of spin-photon coupling in the 
cavity output field, which may provide guidance to the experimental search for strong coupling in such spin-photon systems and 
 opens the way to cavity-based readout of the spin qubit.
\end{abstract}

\pacs{42.50.Pq, 73.21.La, 03.67.Lx, 85.35.Gv}

\maketitle
\section{Introduction}

Building a practical solid state quantum processor necessitates a flexible scheme of coupling individual qubits such that a 2D array of qubits, or even a network with connectivity between arbitrary pairs of qubits (``all-to-all'' connectivity), may be achieved \cite{Fowler2012,Corcoles2015,Debnath2016}. For superconducting qubits, entanglement of qubits separated by macroscopic distances has been demonstrated using the approach of circuit quantum electrodynamics (cQED)~\cite{Raimond2001,Blais2004,Childress2004,2017arXiv170900466C}, whereby photons confined inside microwave frequency cavities serve as mobile carriers of quantum information that mediate long-range qubit interactions \cite{Majer2007,Sillanpaa2007}. Compared to superconducting qubits, qubits based on spins of electrons in semiconductor quantum dots (QDs) have the virtue of long lifetimes ($T_1$) that can be on the order of seconds for Si \cite{Yang2013,Zwanenburg2013}. On the other hand, the coupling of spin qubits has remained limited by nearest neighbor exchange interactions with typical distances $<$100 nm \cite{Petta2005,Veldhorst2015}. The development of a spin-cQED architecture
in which spin qubits are coherently coupled to microwave frequency photons is therefore a critical goal which would enable a spin-based quantum processor with full connectivity.

To transfer quantum states between a spin qubit and a cavity photon with high fidelity, it is necessary to achieve the strong-coupling regime in which the spin-photon coupling  $g_s$ exceeds both the cavity decay rate $\kappa$ and the spin decoherence rate $\gamma_s$ \cite{Raimond2001,Imamoglu2009}. While demonstrations of strong coupling have already been made with superconducting qubits \cite{Wallraff2004} and semiconductor charge qubits \cite{Mi2017b,Stockklauser2017,Bruhat2016}, such a task has proven challenging for a single spin due to its small magnetic dipole, which results in coupling rates that are less than 1 kHz and too slow compared to typical spin dephasing rates \cite{Imamoglu2009,Amsuss2011,Bienfait2016,Eichler2017}. An alternative route toward strong spin-photon coupling involves hybridizing the spin and charge states of QD electrons \cite{Childress2004,Trif2008,Cottet2010,Hu2012,Beaudoin2016}. The relatively large electric susceptibilities of the electron charge states lead to an effective spin-photon coupling rate $g_s$ on the order of MHz, as recently demonstrated by a carbon nanotube double quantum dot (DQD) device \cite{Viennot2015}. However, spin-charge hybridization also renders spin qubits susceptible to charge noise, which has up to now 
prevented the strong coupling regime from being reached with a single spin~\cite{Viennot2015}.

\begin{figure}
\includegraphics[width=1\columnwidth]{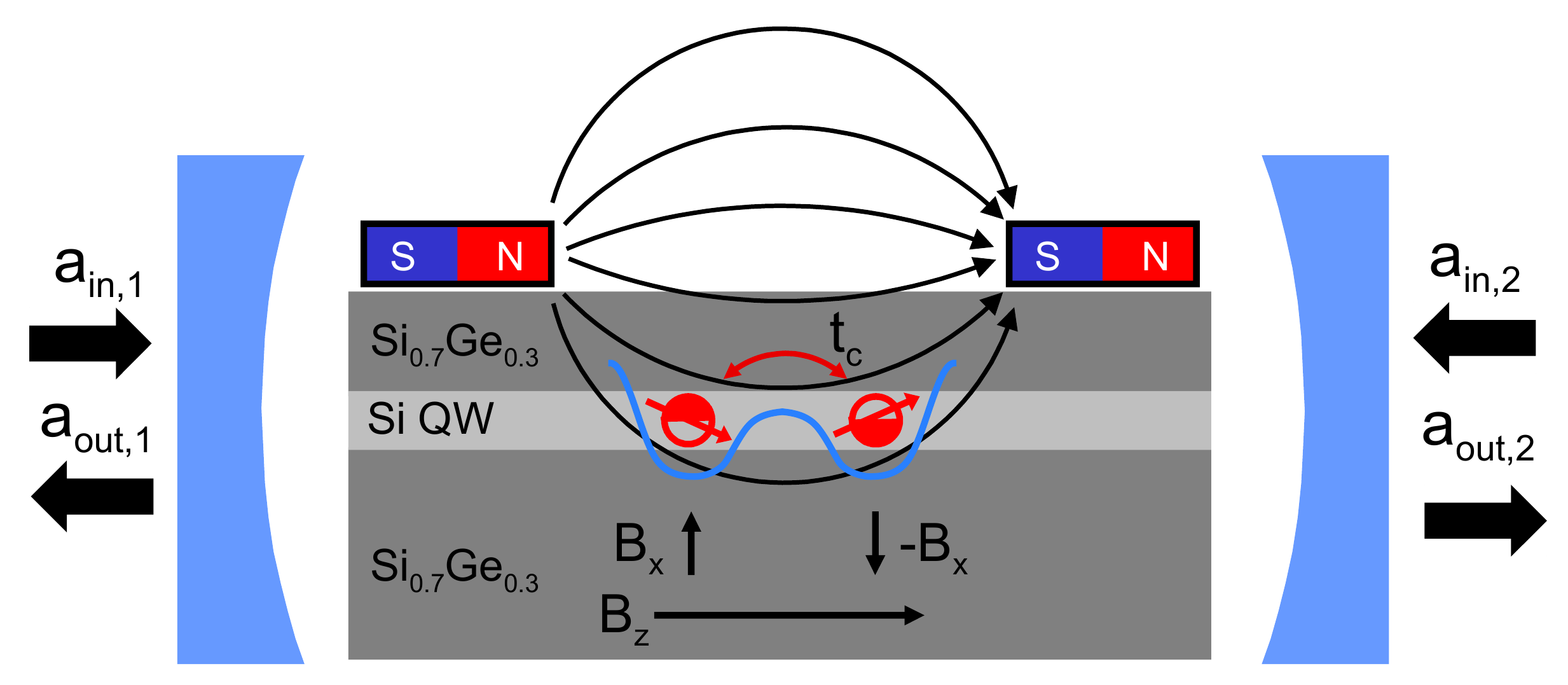}
\protect\caption{\label{fig:figure1-1}  Schematic illustration of the Si gate-defined DQD influenced by an homogeneous external magnetic field, $B_z$, and the inhomogeneous perpendicular magnetic field created by  a micromagnet, with opposite direction at the positions of the two QDs, $\pm B_x$. The DQD is electric-dipole-coupled to the microwave cavity represented in blue. 
The cavity field is excited at the left and right ports via $a_{\mathrm{in},1}$ and $a_{\mathrm{in},2}$, and the output can be measured  either at the left ($a_{\mathrm{out},1}$) or right port ($a_{\mathrm{out},2}$).}
\end{figure} 

Here we analyze a scheme for strong spin-photon coupling using a semiconductor DQD placed in the inhomogeneous magnetic field of a micromagnet, first outlined in Ref.~\cite{Hu2012}. We extend this previous work by predicting a complete map of the effective spin-photon coupling rate $g_s$ and spin decoherence rate $\gamma_s$. 
This allows us to find optimal working points for coherent spin-photon coupling. 
We further present detailed calculations of the cavity transmission and identify experimentally observable signatures of spin-photon coupling. Importantly, we predict that the strong-coupling regime between a single spin and a single photon is achievable in Si using values of the charge-cavity vacuum Rabi frequency $g_c$ and charge decoherence rate $\gamma_c$ from recent experiments \cite{Mi2017,Mi2017b}.

\begin{figure}
\includegraphics[width=1\columnwidth]{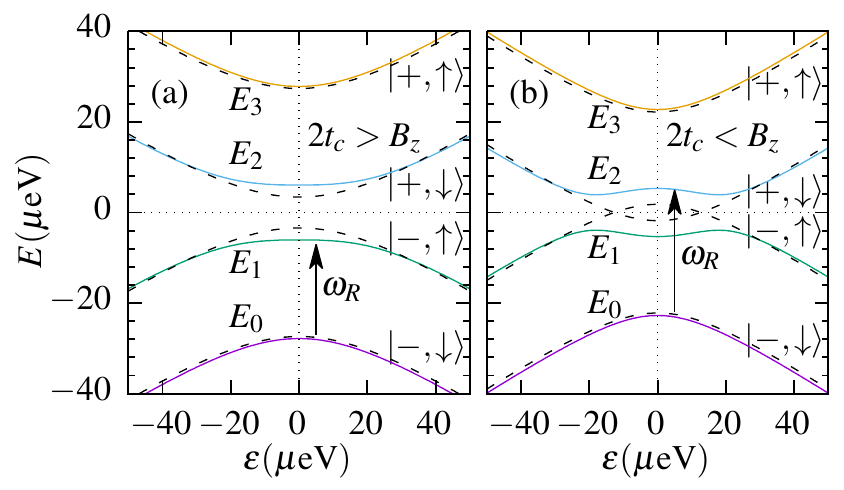}
\includegraphics[width=1\columnwidth]{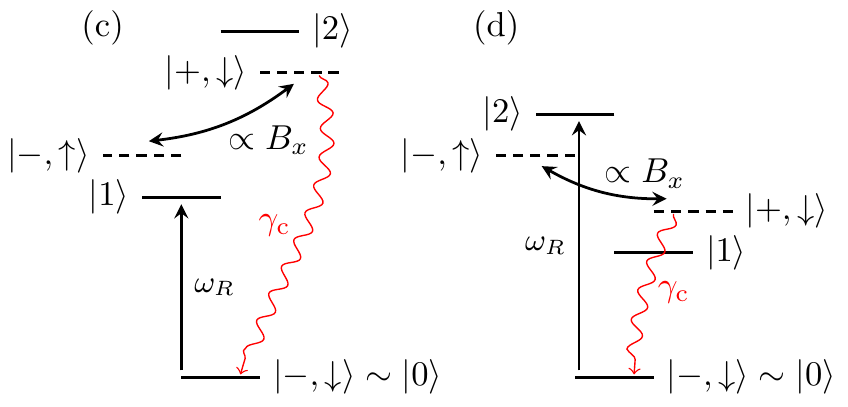}
\protect\caption{\label{fig:figure1}(a,b) Energy levels $E_n\, (n=0,\dots 3)$ as a function of the DQD detuning parameter $\epsilon$. The dashed lines  are the energy levels without a magnetic  field gradient  ($B_x=0$).
They correspond to the bonding (+) and antibonding (-) orbitals with spin $\uparrow,\downarrow$ in the $z$-direction, denoted by $\left|\pm,\uparrow(\downarrow)\right\rangle$.
The arrow represents the transition driven by the probe field, at frequency  $\omega_{\mathrm{R}}$. Here, we choose the parameters $B_z=24\,\mu \mathrm{eV}$ and $B_x=10\,\mu \mathrm{eV}$. For the tunnel coupling: (a) $t_c=15.4\,\mu \mathrm{eV}>B_z/2$ and (b) $t_c=10.2\,\mu \mathrm{eV}<B_z/2$. (c,d) Schematic representation of the $\Lambda$-system that captures the essential dynamics  in (a) and (b), respectively (near $\epsilon=0$). If the orbital energy, $\Omega=\sqrt{\epsilon^2+4t_c^2}$, is near $B_z$, the  levels $\left|-,\uparrow\right\rangle$ and $\left|+,\downarrow\right\rangle$ hybridize into the states $\left|1\right\rangle$ and $\left|2\right\rangle$ due to the magnetic  field gradient, while the ground state is approximately unperturbed $\left|0\right\rangle\sim\left|-,\downarrow\right\rangle$. The wavy lines represent  charge decoherence with rate $\gamma_{\mathrm{c}}$.}
\end{figure} 

The physical system consists of a gate-defined Si DQD that is embedded in a superconducting cavity; see Fig.~\ref{fig:figure1-1}. The electric-dipole interaction couples the electronic charge states in the DQD to the cavity electric field. 
The introduction of an inhomogeneous magnetic field, as sketched in Fig.~\ref{fig:figure1-1}, 
 hybridizes the charge states of a DQD electron with its spin states, indirectly coupling the cavity electric field to the electron spin.

\section{The model}
\label{sec:model}

We assume that the DQD is filled with a single electron and has two charge configurations, with the electron located either on the left (L) or  right (R) dot, with onsite energy difference (detuning) $\epsilon$ and tunnel coupling $t_c$. If  a homogeneous magnetic field $B_z$ and a perpendicular spatial gradient field $B_x$ are applied we can model the single electron DQD with the Hamiltonian  
\begin{equation}
H_{0}  = \frac{1}{2}\left(\epsilon \tau_z+2t_c\tau_x+B_z \sigma_z+B_x \sigma_x \tau_z\right), \label{eq:DQD}
\end{equation}
where $\tau_{\alpha}$ and $\sigma_{\alpha}$ are the Pauli operators in position (L,R) and spin space, respectively. Here, $B_{z(x)}$ are the magnetic fields in energy units and $\hbar=1$. This is a 4-level Hamiltonian with eigenenergies $E_n$ and eigenstates $\left|n\right\rangle$ for $n=0,..,3$.
The eigenenergies in the regime $2t_c>B_z$ ($2t_c<B_z$) are shown in 
Fig.~\ref{fig:figure1}~(a) [Fig.~\ref{fig:figure1}~(b)].
The magnetic field gradient generates spin-charge hybridization,  coupling the original ($B_x=0$) energy levels  (dashed lines) and   inducing  anticrossings at $\epsilon=\pm\sqrt{4t_c^2+B_z^2}$ if $2t_c<B_z$; see Fig~\ref{fig:figure1}~(b).

In the dipole approximation, the coupling of the DQD to the electric field of a microwave cavity can be described as
\begin{equation}
H_{\mathrm{I}}  =g_c\left(a+a^{\dagger}\right)\tau_z\  , \label{eq:cavity-DQD}
\end{equation}
where $a$ and $a^{\dagger}$ are the bosonic cavity photon creation and annihilation  operators. The Hamiltonian for the relevant cavity mode, with frequency $\omega_{\mathrm{c}}$ is $H_{\mathrm{c}}=\omega_{\mathrm{c}} a^{\dagger}a$. 
In the eigenbasis of $H_{0}$, 
the interaction  acquires non-diagonal elements, 
\begin{equation}
H_{\mathrm{I}}  = g_c\left(a+a^{\dagger}\right)\sum_{n,m=0}^{3}d_{nm}\left|n\right\rangle\left\langle m\right|\ . \label{eq:cavity-DQD-eigenbasis}
\end{equation} 

As we will show below, the essential dynamics of this system can also be described in terms of a so called $\Lambda$-system, with two weakly-coupled excited states and a ground state; see Figs.~\ref{fig:figure1}~(c) and~(d)~\cite{Palmer2015,Kubo,Jacques2016}.

\section{Input-output theory}
\label{sec:input-output}

To treat the DQD and the cavity as an open system, we move into the Heisenberg picture and use the  quantum Langevin equations (QLEs) for the system operators, including the photon operators $a,a^{\dagger},$ and $\sigma_{nm}=\left|n\right\rangle\left\langle m\right|$.  This treatment enables the calculation of the outgoing fields, $a_{\mathrm{out},1}$ and $a_{\mathrm{out},2}$, at the two  cavity ports given the incoming weak fields, $a_{\mathrm{in},1}$ and $a_{\mathrm{in},2}$~\citep{Collett1984,Gardiner1985,Burkard2016}.

If the average population of the energy levels, $p_n\equiv\left\langle \sigma_{nn}\right\rangle$,  follows a thermal distribution, the linear response to a probe field is reflected in the dynamics of the non-diagonal operators $\sigma_{nm}$. 
If the cavity is driven with a microwave field with a near-resonant frequency $\omega_{\mathrm{R}}$, the QLEs in a frame rotating with the driving frequency read
\begin{eqnarray}
\dot{a} & =&i \Delta_0 a-\frac{\kappa}{2} a+\sqrt{\kappa_1}a_{\mathrm{in},1}+\sqrt{\kappa_2}a_{\mathrm{in},2}\label{eq:LLangevin1}\\
&& -i g_c e^{i\omega_{R}t} \sum_{n,m=0}^{3}d_{nm}\sigma_{nm},\nonumber\\
\dot{\sigma}_{nm} & =&-i \left(E_m-E_n\right) \sigma_{nm}-\sum_{n'm'}\gamma_{nm,n'm'} \sigma_{n'm'}  \label{eq:LLangevin3}\\
&&+\sqrt{2\gamma}\mathcal{F}  -i g_c \left(a e^{-i\omega_{R}t}+a^{\dagger} e^{i\omega_{R}t} \right)d_{mn}(p_n-p_m), \nonumber
\end{eqnarray} 
where $\Delta_0=\omega_{\mathrm{R}}-\omega_{\mathrm{c}}$ is the   detuning of the driving field relative to the cavity frequency, $\kappa$ is the total cavity decay rate, with $\kappa_{1,2}$ the decay rates through the input and output ports. 
$\mathcal{F}$  is the quantum noise of the DQD and $a_{\mathrm{in},i}$ denote the incoming parts of the external field at the ports. The outgoing fields can be calculated as $a_{\mathrm{out},i}=\sqrt{\kappa_i}a-a_{\mathrm{in},i}$.
The superoperator  $\gamma$, with matrix elements $\gamma_{nm,n'm'}$, represents the decoherence  processes which, in general, can couple  the equations for the operators $\sigma_{nm}$. 
In this work, the decoherence superoperator $\gamma$ will capture charge relaxation and dephasing due to charge noise (see Appendix~\ref{sec:decoherence}), since these are the most relevant sources of decoherence.

This formalism allows us to compute the transmission through the microwave cavity.
Within a rotating-wave approximation (RWA) (see Appendix~\ref{sec:RWA}) we can eliminate the explicit time-dependence in Eqs.~\eqref{eq:LLangevin1} and~\eqref{eq:LLangevin3} and solve the equations for the expected value of these operators in the stationary limit ($\bar{a}$, $\bar{\sigma}_{n,m}$) to obtain the susceptibilities,
\begin{equation}
\bar{\sigma}_{n,n+j}  =\chi_{n,n+j}\bar{a}; \quad (j=1,...,3-n) , \label{eq:suscnm}
\end{equation}
and the transmission  ${A=\bar{a}_{\mathrm{out},2}/\bar{a}_{\mathrm{in},1}}$,
\begin{equation}
A  = \frac{-i\sqrt{\kappa_1 \kappa_2}}{-\Delta_0-i\kappa/2+g_c \sum_{n=0}^{2} \sum_{j=1}^{3-n}d_{n,n+j}\chi_{n,n+j}}, \label{eq:suscnm}
\end{equation}
which is in general a complex quantity. 
We have considered here  $\left\langle a_{\mathrm{in},2}\right\rangle=0$ and $\left\langle\mathcal{F}\right\rangle=0$.

\section{Orbital basis}
\label{sec:orbital-basis}

In the product basis of antibonding and  bonding orbitals $\pm$ with spin $\uparrow\downarrow$ in the $z$-direction,  $\{\left|+,\uparrow\right\rangle,{\left|-,\uparrow\right\rangle},\left|+,\downarrow\right\rangle,\left|-,\downarrow\right\rangle\}$, the Hamiltonian in Eq.~\eqref{eq:DQD} reads
\begin{align}
H_{0}^{\mathrm{orb}}&=\frac{1}{2}\begin{pmatrix}
\Omega+B_z & 0 &  B_x\sin{\theta} & - B_x\cos{\theta} \\
0 & -\Omega+B_z & - B_x\cos{\theta} &  - B_x\sin{\theta}  \\
 B_x\sin{\theta} & - B_x\cos{\theta} &  \Omega-B_z & 0 \\
- B_x\cos{\theta} & -B_x\sin{\theta} & 0 & -\Omega-B_z
\end{pmatrix}, \label{eq:Horbital-basis}
\end{align}
where $\Omega=\sqrt{\epsilon^2+4t_c^2}$ is the orbital energy and we introduce ${\theta=\arctan{\frac{\epsilon}{2t_c}}}$ as the ``orbital angle''. In this basis the dipole operator takes the form
\begin{equation}
d^{\mathrm{orb}}=\begin{pmatrix}
\sin{\theta} & -\cos{\theta} & 0 & 0 \\
-\cos{\theta} & -\sin{\theta} & 0 & 0  \\
0 & 0 & \sin{\theta} & -\cos{\theta} \\
0 & 0 & -\cos{\theta}& -\sin{\theta}
\end{pmatrix}
.\label{eq:dmatrix}
\end{equation}
In the simplest case, $\epsilon=0$,  the orbital angle $\theta$ is zero, and we can rewrite the Hamiltonian as
\begin{equation}
H_{0}^{\mathrm{orb}}(\epsilon=0)=\frac{r}{2}\begin{pmatrix}
\frac{2t_c+B_z}{r} & 0 &  0 & - \sin{\Phi} \\
0 & - \cos{\Phi} & - \sin{\Phi} & 0 \\
 0 & - \sin{\Phi} &  \cos{\Phi} & 0 \\
- \sin{\Phi} & 0 & 0 & \frac{-2t_c-B_z}{r}
\end{pmatrix}
,  \label{eq:Horbital-basis-eps0}
\end{equation}
with ${r=\sqrt{(2t_c-B_z)^2+B_x^2}}$ and the spin-orbit mixing angle ${\Phi=\arctan{\frac{B_x}{2t_c-B_z}}}$ ($\Phi\in (0,\pi)$).
As the dipole operator couples the states $\left|-,\downarrow\right\rangle$ and $\left|+,\downarrow\right\rangle$  and the field gradient couples  $\left|+,\downarrow\right\rangle$ to  $\left|-,\uparrow\right\rangle$, the combination of these two effects leads to a coupling between the two different spin states $\left|-,\downarrow\right\rangle$ and $\left|-,\uparrow\right\rangle$. 
It is this coupling that can be harnessed to 
coherently hybridize a single electron spin with a single photon and 
achieve the strong-coupling regime.

\section{Results}
\label{sec:Results}

\subsection{Effective coupling at zero detuning}
\label{subsec:coupling-eps0}

The spin-charge hybridization created by the inhomogeneous magnetic field
allows for the coupling of the spin to the cavity. This is visible in the form of the operator $d$ in the eigenbasis; see Eq.~\eqref{eq:cavity-DQD-eigenbasis}.
In the simple case of zero DQD detuning, $\epsilon=0$, the ordered energy levels are
\begin{eqnarray}
E_{3,0} & =&\pm\frac{1}{2}\sqrt{(2t_c+B_z)^2+B_x^2} ,\\
E_{2,1} & =&\pm\frac{1}{2}\sqrt{(2t_c-B_z)^2+B_x^2} . \label{eq:energies}
\end{eqnarray} 
Using the spin-orbit mixing angle $\Phi$, the eigenstates $\left|1\right\rangle$ and $\left|2\right\rangle$ can be expressed as
\begin{eqnarray}
\left|1\right\rangle&= & \cos{\frac{\Phi}{2}}\left|-,\uparrow\right\rangle+\sin{\frac{\Phi}{2}}\left|+,\downarrow\right\rangle, \label{eq:vectors1}\\
\left|2\right\rangle&= & \sin{\frac{\Phi}{2}}\left|-,\uparrow\right\rangle-\cos{\frac{\Phi}{2}}\left|+,\downarrow\right\rangle, \label{eq:vectors2}
\end{eqnarray}
while the other two can be approximated by 
\begin{eqnarray}
\left|0\right\rangle&\simeq & \left|-,\downarrow\right\rangle, \label{eq:vectors3}\\
\left|3\right\rangle&\simeq & \left|+,\uparrow\right\rangle, \label{eq:vectors4}
\end{eqnarray}
if $r\ll (2t_c+B_z)$, i.e., for small $|2t_c-B_z|$.
In this limit, the dipole matrix elements,
\begin{equation}
d=\begin{pmatrix}
0 & d_{01} & d_{02} & 0 \\
d_{01} & 0 & 0 & d_{13}  \\
d_{02} & 0 & 0 & d_{23} \\
0 & d_{13} & d_{23}& 0
\end{pmatrix}
,\label{eq:dmatrix}
\end{equation}
simplify to
\begin{eqnarray}
d_{01}&=&d_{23}  \simeq -\sin\frac{\Phi}{2},\\
d_{02}&=&-d_{13}  \simeq \cos\frac{\Phi}{2}. \label{eq:delements}
\end{eqnarray}
This means that the hybridization due to the  weak magnetic field gradient generates an 
effective coupling between the levels $\left|-,\downarrow\right\rangle$ and $\left|-,\uparrow\right\rangle$, with opposite spin. 
The spin nature of the transitions $0\leftrightarrow 1$ and $0\leftrightarrow 2$ depends on the spin-orbit mixing angle $\Phi$; see Eqs.~\eqref{eq:vectors1} to~\eqref{eq:vectors3}. 
Assuming that the cavity frequency is tuned to the predominantly spin-like transition, 
which is $0\leftrightarrow1$ ($0\leftrightarrow2$) for $\cos\Phi>0$ ($\cos\Phi<0$), the effective spin-cavity coupling strength will be given by $g_{\mathrm{s}}=g_c|d_{01(2)}|$.

\subsection{Effective coupling at $\epsilon\neq 0$}
\label{subsec:coupling-epsno0}

For $\epsilon\neq 0$ the energy levels are
\begin{align}
E_{3,0}  &={\pm}\frac{1}{2}\left[\left(\Omega+\sqrt{B_z^2+B_x^2 \sin^2{\theta}}\right)^2{+}B_x^2\cos^2{\theta}\right]^{1/2} ,\\
E_{2,1}  &={\pm}\frac{1}{2}\left[\left(\Omega-\sqrt{B_z^2+B_x^2 \sin^2{\theta}}\right)^2{+}B_x^2\cos^2{\theta}\right]^{1/2} .\label{eq:energies-general}
\end{align}
Analogously to the previous section,
if
$\sqrt{(\Omega-B_z)^2+B_x^2}\ll (\Omega+B_z)$
we can approximate the eigenstates by Eqs.~\eqref{eq:vectors1} to~\eqref{eq:vectors4}
where the spin-orbit mixing angle is now $\Phi=\arctan{\frac{B_x \cos\theta}{\Omega-B_z}}$  ($\Phi\in (0,\pi)$).
Within this approximation,
\begin{eqnarray}
d_{01}&=&d_{23}  \simeq-\cos\theta\sin\frac{\Phi}{2}, \label{eq:delements2}\\
d_{02}&=&-d_{13}  \simeq \cos\theta\cos\frac{\Phi}{2}. \label{eq:delements3}
\end{eqnarray}

\subsection{Effective coupling map}
\label{subsec:coupling-map}

\begin{figure}
\includegraphics[width=\columnwidth]{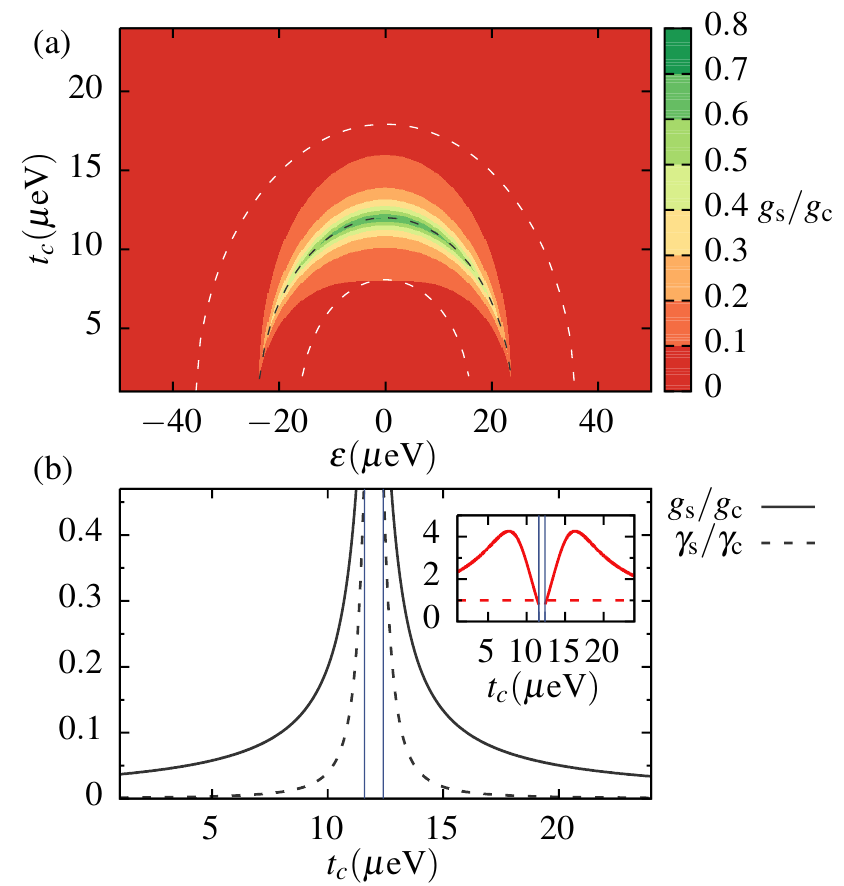}
\protect\caption{\label{fig:figure3}(a) Expected effective coupling $g_{\mathrm{s}}/g_{\mathrm{c}}=|d_{01(2)}|$, according to Eqs.~\eqref{eq:delements2} and~\eqref{eq:delements3}  as a function of $t_c$ and $\epsilon$. The black dashed line corresponds to $\Omega=B_z$.
The most interesting region lies in between the two white dashed lines, where our approximations are  accurate ($\sqrt{(\Omega-B_z)^2+B_x^2}\ll (\Omega+B_z)$).
We chose  $B_x=1.62\,\mu\mathrm{eV}$ and $B_z=24\,\mu\mathrm{eV}$. (b) Spin-photon coupling strength $g_{\mathrm{s}}/g_{\mathrm{c}}$ and spin decoherence rate $\gamma_{\mathrm{s}}/\gamma_{\mathrm{c}}$ as a function of  $t_c$ for $\epsilon=0$, $B_z=B_z^{\mathrm{res}}$, and $B_x=1.62\,\mu\mathrm{eV}$. Between the two blue vertical lines, the resonance cannot be achieved by tuning $B_z$. Inset: ratio $g_{\mathrm{s}}/\sqrt{(\gamma_{\mathrm{s}}^2+(\kappa/2)^2)/2}$ in the same range. The coupling is strong when this quantity is larger than one (dashed line). We have chosen   $\gamma_c/2\pi=100\,\mathrm{MHz}$, $g_c/2\pi=40\,\mathrm{MHz}$, and $\kappa/2\pi=1.77\,\mathrm{MHz}$.}
\end{figure}

Before calculating the effect of the Si DQD on the cavity transmission $A$, let us estimate the magnitude of the coupling $g_{\mathrm{s}}$.
For $\Omega>B_z$ ($\Omega<B_z$),  $0\leftrightarrow 1$ ($0\leftrightarrow 2$) is predominantly a spin transition, 
therefore
we can obtain a map for the effective  coupling by using $g_{\mathrm{s}}=g_c|d_{01(2)}|$; see  Fig.~\ref{fig:figure3}~(a). As  the value of $\Omega$ approaches $B_z$, $\Phi$ tends to $ \pi/2$ and  the  coupling  is maximized.
However, in this regime, due to strong spin-charge hybridization, 
the charge nature of the transition increases [see Eqs.~\eqref{eq:vectors1} and~\eqref{eq:vectors2}] 
and with it the decoherence rate increases, preventing the system from 
reaching strong coupling.
In the following we show that
the ratio of the coupling rate to the total 
decoherence rate can be optimized by working away from maximal coupling. In particular the strong-coupling regime for the spin can be achieved.

\subsection{Cavity transmission}
\label{subsec:trasmission}

In the following, we consider the DQD to be in its ground state, such that $p_n=\delta_{n,0}$ in Eq.~\eqref{eq:LLangevin3}. If the cavity frequency is close to the Zeeman energy, $\omega_{\mathrm{c}}\sim B_z$, the transition $0\leftrightarrow 3$ is off-resonant and the relevant dynamics is contained in the level structure of Figs.~\ref{fig:figure1}~(c) and~(d). Moreover, this transition is not coupled to the others since $d_{03}=0$ and $\gamma_{03,nm}=\delta_{n0}\delta_{m3}$ (see Appendix~\ref{sec:decoherence}). To calculate  the cavity response, it is sufficient to solve the  QLEs for $\left\langle a\right\rangle$, $\left\langle\sigma_{01}\right\rangle$ and $\left\langle\sigma_{02}\right\rangle$ (in the following we omit the brackets) within the RWA (see Appendix~\ref{sec:RWA}).

As explained above, the decoherence processes accounted for in Eq.~\eqref{eq:LLangevin3} can result in  a different decay rate for every transition  and can also couple different  transitions.
As shown in Appendix~\ref{sec:decoherence}, the decoherence superoperator in the basis $\{\sigma_{01} ,\sigma_{02} \}$ reads
\begin{eqnarray}\gamma=
\gamma_c\begin{pmatrix}
\sin^2\frac{\Phi}{2} &  -\frac{\sin{\Phi}}{2} \\
-\frac{\sin{\Phi}}{2} & \cos^2\frac{\Phi}{2} 
\end{pmatrix}
,\label{eq:dmatrix}
\end{eqnarray} 
where $\gamma_c=\gamma_1/2+\gamma_{\phi}$ contains charge relaxation, $\gamma_1$, and pure charge dephasing, $\gamma_{\phi}$. 
With this, the QLEs read
\begin{eqnarray}
\dot{a}  &=& i \Delta_0 a-\frac{\kappa}{2} a+\sqrt{\kappa_1} a_{\mathrm{in},1}\nonumber\\
&& -i g_c ( d_{01}\sigma_{01}+ d_{02}\sigma_{02}), \label{eq:a}\\
\dot{\sigma}_{01} & =&-i \delta_1 \sigma_{01}-\gamma_c \sin^2\frac{\Phi}{2} \sigma_{01} \ \nonumber\\
&& +\frac{\gamma_c}{2} \sin{\Phi}\sigma_{02} -i g_c a d_{10}, \label{eq:01}\\
\dot{\sigma}_{02} & =&-i \delta_2 \sigma_{02}-\gamma_c \cos^2\frac{\Phi}{2} \sigma_{02} \nonumber\\
& &+\frac{\gamma_c}{2} \sin{\Phi}\sigma_{01} -i g_c a d_{20}
\ ,\label{eq:02}
\end{eqnarray} 
with the detunings $\delta_n\equiv E_n-E_0-\omega_{\mathrm{R}}$ ($n=1,2$). The solution of these equations in the stationary limit  allows us to  compute   the susceptibilities 
\begin{eqnarray}
\chi_{01} &=&\frac{\bar{\sigma}_{01}}{\bar{a}}=\frac{g_c \cos\theta \sin(\Phi/2)}{\delta_1-i \gamma_{\mathrm{eff}}^{(2)}},\\
\chi_{02} &=&\frac{\bar{\sigma}_{02}}{\bar{a}}=\frac{-g_c \cos\theta\cos(\Phi/2)}{\delta_2-i \gamma_{\mathrm{eff}}^{(1)}}, \label{eq:susc0102}
\end{eqnarray}
where $\gamma_{\mathrm{eff}}^{(n)}\equiv \gamma_c [\delta_2 \sin^2(\Phi/2)+\delta_1 \cos^2(\Phi/2)]/\delta_n$, and the transmission through the cavity  
\begin{align}
A  &= \frac{-i\sqrt{\kappa_1 \kappa_2}}{-\Delta_0-i\frac{\kappa}{2}+g_c \left(\chi_{01}d_{01}+\chi_{02}d_{02}\right)}, \label{eq:afinal}
\end{align}
with $d_{01}$ and $d_{02}$ defined in Eqs.~\eqref{eq:delements2} and~\eqref{eq:delements3}.
If  $0\leftrightarrow 1$ ($0\leftrightarrow 2$) is predominantly a spin transition and the corresponding transition energy  is in resonance with the cavity frequency, we expect an effective spin decoherence rate $\gamma_{\mathrm{s}}=\gamma_{\mathrm{eff}}^{(2)}$ ($\gamma_{\mathrm{s}}=\gamma_{\mathrm{eff}}^{(1)}$). 
In Fig.~\ref{fig:figure3}~(b) we show the ratio $\gamma_{\mathrm{s}}/\gamma_{\mathrm{c}}$, together with $g_{\mathrm{s}}/g_{\mathrm{c}}$, as a function of the tunnel coupling for $\epsilon=0$. Here, we have set the  external  magnetic field to the resonant value $B_z^{\mathrm{res}}$  such that $E_{1(2)}-E_0=\omega_{\mathrm{c}}$. This is
\begin{equation}
B_z^{\mathrm{res}}=\omega_{\mathrm{c}}\sqrt{1-\frac{B_x^2}{\omega_{\mathrm{c}}^2-4t_c^2}}, \label{eq:resonant-field}
\end{equation} 
for $\epsilon=0$. In a small region around $2t_c\sim \omega_{\mathrm{c}}$ ($\sqrt{\omega_{\mathrm{c}}(\omega_{\mathrm{c}}-B_x)}<2t_c<\sqrt{\omega_{\mathrm{c}}(\omega_{\mathrm{c}}+B_x)}$), indicated with the vertical lines in Fig.~\ref{fig:figure3}~(b), it is not possible to achieve the desired resonance by tuning $B_z$. 
We observe that in the wings  of the peak $g_{\mathrm{s}}/g_{\mathrm{c}}\gg \gamma_{\mathrm{s}}/\gamma_{\mathrm{c}}$, which may lead the spin-cavity system to be in the strong-coupling regime even when the charge-cavity system is not ($g_{\mathrm{c}}<\gamma_{\mathrm{c}}$).
This is visible  in the inset, where we show that the ratio $g_{\mathrm{s}}/\sqrt{(\gamma_{\mathrm{s}}^2+(\kappa/2)^2)/2}$  exceeds one,  signifying the strong-coupling regime (see Appendix~\ref{sec:SC}).
 
\begin{figure}
\includegraphics[width=\columnwidth]{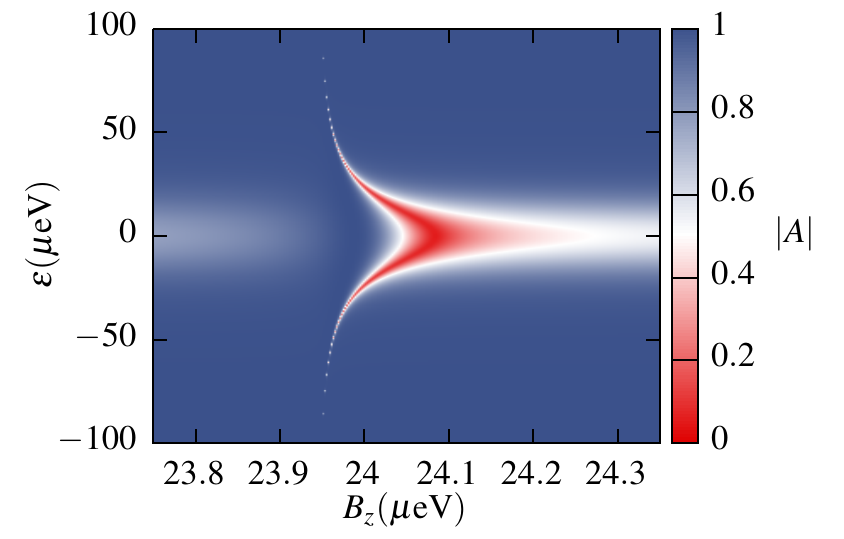}
\protect\caption{\label{fig:figure4}Cavity transmission spectrum, $|A|$, as a function of $B_z$ and $\epsilon$ at zero detuning $\Delta_0=0$. The other parameters are $t_c=15.4\,\mu\mathrm{eV}$, $B_x=1.62\,\mu\mathrm{eV}$, $\gamma_1/2\pi=200\,\mathrm{MHz}$, $\gamma_{\phi}/2\pi=150|\sin{\theta}|\,\mathrm{MHz}$, $g_c/2\pi=40\,\mathrm{MHz}$, $\kappa/2\pi=1.77\,\mathrm{MHz}$, and $\omega_{\mathrm{c}}/2\pi=5.85\,\mathrm{GHz}\simeq 24\,\mu\mathrm{eV}$.}
\end{figure}

\begin{figure}
\includegraphics[width=\columnwidth]{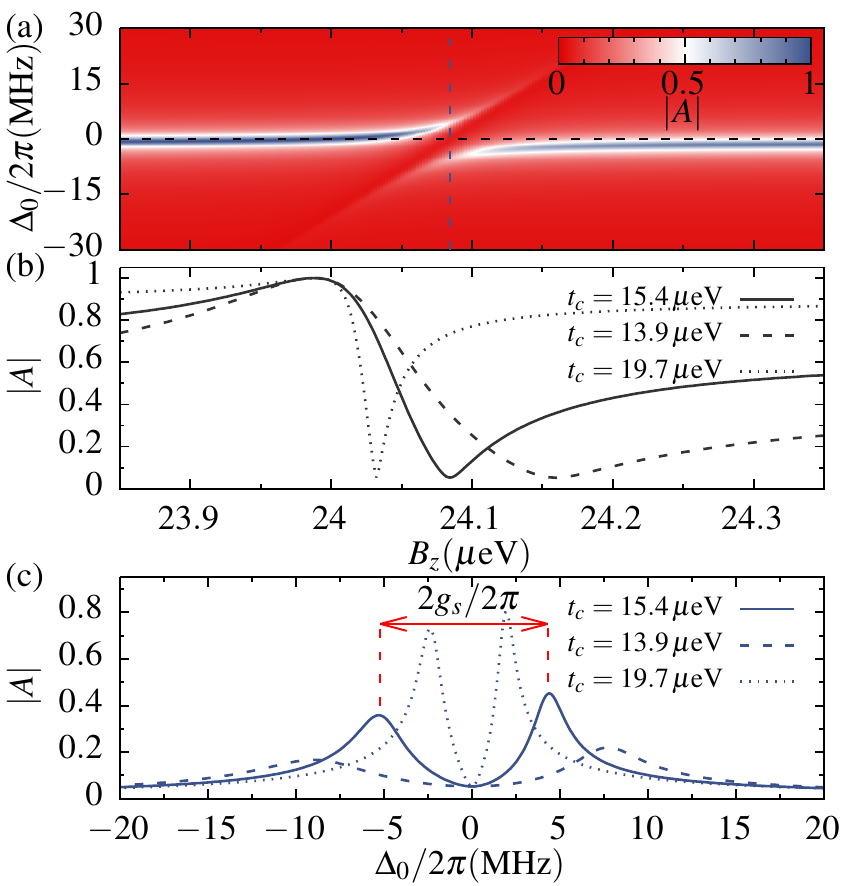}
\protect\caption{\label{fig:figure2}(a) Cavity transmission spectrum $|A|$ as a function of $B_z$ and $\Delta_0$. (b) ((c)) shows $|A|$ as a function of $B_z$ ($\Delta_0$) for the value of $\Delta_0$ ($B_z$) indicated by the black (blue) dashed line.
In (a)  $t_c=15.4\,\mu\mathrm{eV}$, while in (b) and (c) we show the result for this value (solid line) and for $t_c=13.9\,\mu\mathrm{eV}$ (dashed line) and $t_c=19.7\,\mu\mathrm{eV}$ (dotted line).  The other parameters are $\epsilon=0$, $B_x=1.62\,\mu\mathrm{eV}$, $\gamma_c/2\pi=100\,\mathrm{MHz}$, $g_c/2\pi=40\,\mathrm{MHz}$, $\kappa/2\pi=1.77\,\mathrm{MHz}$, and $\omega_{\mathrm{c}}/2\pi=5.85\,\mathrm{GHz}\simeq 24\,\mu\mathrm{eV}$.}
\end{figure}

According to the level structure, we expect to observe a signature of spin-photon coupling by driving the cavity near resonance ($\Delta_0\sim0$) and swiping the external magnetic field  through the cavity frequency.
In Fig.~\ref{fig:figure4} we show the calculated transmission through the cavity 
    as a function of the external magnetic field $B_z$ and the DQD detuning $\epsilon$ when the driving frequency matches the cavity frequency. We have chosen $\kappa_1=\kappa_2=\kappa/2$. When the cavity frequency is close to the transition energy $0\leftrightarrow 1$, the interaction between the electron and the cavity field   
 results in a significantly reduced cavity transmission.
 Interestingly, close to $B_z\sim\omega_{\mathrm{c}}$ the transmission approaches one due to an interference between the two energy levels. At this point, $\chi_{01}d_{01}+\chi_{02}d_{02}\simeq0$.

In the usual scenario of a two-level system 
coupled to a photonic cavity, 
strong coupling results in light-matter hybridization, as evidenced
 in the observation of vacuum Rabi splitting 
in the cavity transmission spectrum when the qubit transition frequency matches the cavity frequency.
The two vacuum Rabi normal modes are separated by a frequency corresponding to the characteristic rate of the light-matter interaction, and the linewidth of each mode reflects the  average decoherence rate of light and matter~\cite{Thompson1992}.
In  Fig.~\ref{fig:figure2}~(a), we show
the absolute value of the transmission, $|A|$, as a function of the magnetic field $B_z$ and the driving frequency relative to $\omega_{\mathrm{c}}$, $\Delta_0$,  at $\epsilon=0$. The phase gives similar information (not shown). 
When the driving frequency is near the cavity frequency, $\Delta_0\sim0$, two peaks emerge in the cavity transmission, signifying the strong-coupling regime.
In Figs.~\ref{fig:figure2}~(b) and~(c) we show the horizontal and vertical cuts of this figure at $\Delta_0=0$ and $B_z=B_z^{\mathrm{res}}$, respectively, where $B_z^{\mathrm{res}}$, given by Eq.~\eqref{eq:resonant-field}, ensures $E_1-E_0=\omega_{\mathrm{c}}$.  In Fig.~\ref{fig:figure2}~(b) we  observe the same interference effect seen in Fig.~\ref{fig:figure4}, and in Fig.~\ref{fig:figure2}~(c) the vacuum Rabi splitting. As indicated  with a red arrow, the effective coupling, related to the separation between the two peaks,  corresponds to $g_{\mathrm{s}}/2\pi\sim 5 \,\mathrm{MHz}$ and the parameters under consideration can be readily achieved in Si DQD architectures \citep{Mi2017,Mi2017b}.

\begin{figure}
\includegraphics[width=\columnwidth]{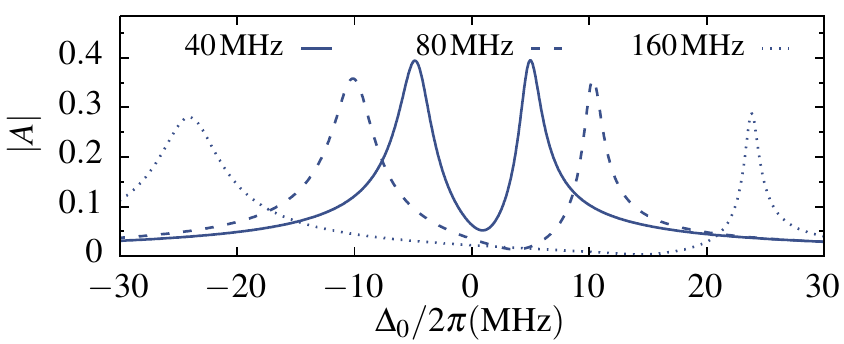}
\protect\caption{\label{fig:figure6}Cavity transmission $|A|$ as a function of $\Delta_0$ close to the resonant field for different values of the charge-cavity coupling  $g_c/2\pi=\{40,80,160\}\,\mathrm{MHz}$. The magnetic field has been slightly detuned from the resonance condition to make the relative heights of the vacuum Rabi split modes the same.  The rest of parameters are  $t_c=15.4\,\mu\mathrm{eV}$, $\epsilon=0$, $B_x=1.62\,\mu\mathrm{eV}$, $\gamma_c/2\pi=100\,\mathrm{MHz}$, $g_c/2\pi=40\,\mathrm{MHz}$, $\kappa/2\pi=1.77\,\mathrm{MHz}$, and $\omega_{\mathrm{c}}/2\pi=5.85\,\mathrm{GHz}\simeq 24\mu\mathrm{eV}$.}
\end{figure}

In the present case, we are dealing with a three-level system, where the spin-photon coupling is mediated by the spin-charge hybridization~\cite{Childress2004}. 
 The three-level system structure explains not only the interference but also why the width and position of the two resonance peaks in Fig.~\ref{fig:figure2}~(c) is slightly asymmetric. As expected, this asymmetry is more apparent as $g_c$ increases, which is shown in Fig.~\ref{fig:figure6}.  In   section~\ref{subsec:two-level-equivalent}, we reduce the problem to an equivalent two-level system to be able to characterize the spin-photon coupling within the standard formalism utilized for the Jaynes-Cummings model.

\subsection{Broadening due to nuclear spins}
\label{subsec:nuclear-spins}

The estimated value of the spin decoherence rate induced by the spin-charge hybridization is of the order $\gamma_{\mathrm{s}}/2\pi\sim 1-10\,\mathrm{MHz}$. Another source of decoherence in Si QDs is the effect of the $^{29}\mathrm{Si}$ nuclear spins which surround the electron spin. As their evolution is slow compared to the typical time scale of the electronic processes, the nuclear spins effectively produce a random magnetic field which slightly influences the total magnetic field on the DQD.
This small perturbation of the magnetic field, $B_z^{\mathrm{tot}}=B_z+B_z^{\mathrm{nuc}}$, will modify the frequency $\omega_{\mathrm{c}}+\Delta_0$ of the two vacuum Rabi normal modes, as can be extracted from Fig.~\ref{fig:figure2}~(a). The nuclear magnetic field follows a Gaussian distribution with average zero and standard deviation $\sigma^{\mathrm{nuc}}$. In this way, a Gaussian profile  is superimposed to the Lorentzian profile of the resonances and the final width also depends on $\sigma^{\mathrm{nuc}}$. At  the point with maximum spin-charge hybridization ($\Omega\sim B_z$), this effect is negligible because the decoherence is dominated by charge decoherence. Away from this point, the two broadening mechanisms have to be combined, resulting in a Voigt profile~\cite{MR2723248}. 
The spin dephasing times in natural Si are $\sim 1\,\mu \mathrm{s}$, which corresponds  approximately to a standard deviation of $\sigma^{\mathrm{nuc}}/2\pi\sim 0.3\,\mathrm{MHz}$ for the  nuclear magnetic field distribution~\cite{Kawakami2014}. 
According to Eq.~\eqref{eq:afinal}, the positions of the two vacuum Rabi modes are given by  the solutions of the equation
\begin{equation}
-\Delta_0+g_{\mathrm{c}}\,\mathrm{Re}\left(\chi_{01}d_{01}+\chi_{02}d_{02}\right)=0 ,\label{eq:Repart}
\end{equation}
where the  susceptibilities are a function of $\Delta_0$ via the detunings $\delta_{1(2)}\equiv E_{1(2)}-E_0-\omega_{\mathrm{R}}$ and $\omega_R=\omega_{\mathrm{c}}+\Delta_0$. As the  magnetic field created by the nuclear spins is small, we can expand the solutions $\Delta_0^{\pm}$ to first order to obtain
\begin{equation}
\Delta_0^{\pm}\simeq\Delta_0^{\pm} (B_z^{\mathrm{nuc}}=0)+\left.\frac{\partial\Delta_0^{\pm}}{\partial B_z}\right|_{B_z^{\mathrm{nuc}}=0}B_z^{\mathrm{nuc}} .
\end{equation}
Therefore the broadening of the vacuum Rabi modes due to the nuclear spins is given by $\sigma=\left|\partial\Delta_0^{\pm}/\partial B_z\right|\sigma^{\mathrm{nuc}}$, and the total spin decoherence rate is
\begin{equation}
\gamma_{\mathrm{s}}^{\mathrm{tot}}=\gamma_{\mathrm{s}}/2+\sqrt{(\gamma_{\mathrm{s}}/2)^2+8(\ln{2})\sigma^2}
.\end{equation}
The long spin dephasing times in Si allow the strong-coupling regime to be reached approximately at the same working points. For instance, for a tunnel coupling $t_c\sim15\,\mu\mathrm{eV}$, the estimated spin dephasing rate induced by charge hybridization is $\gamma_{\mathrm{s}}/2\pi\sim 2\,\mathrm{MHz}$ and the broadening due to the nuclear spins is given by $\sigma/2\pi\sim 0.14\,\mathrm{MHz}$, therefore $\gamma_{\mathrm{s}}^{\mathrm{tot}}/2\pi\sim 2\,\mathrm{MHz}$.

\subsection{Two-level equivalent system}
\label{subsec:two-level-equivalent}

To reduce the problem to a two-level system, it is more convenient to work in the orbital basis. 
Using the relations in Eqs.~\eqref{eq:01abbasis} and~\eqref{eq:02abbasis} we can rewrite the QLEs
in terms of the operators $a$,
$\sigma_{\tau}=\left|-,\downarrow \right\rangle\left\langle +,\downarrow \right|$ and  $\sigma_{s}=\left|-,\downarrow \right\rangle\left\langle -,\uparrow \right|$. Neglecting input noise terms for the charge relaxation which will be irrelevant for our linear response theory, these equations read
\begin{eqnarray}
\dot{a}  &=&i \Delta_0 a-\frac{\kappa}{2}  a+\sqrt{\kappa_1} a_{\mathrm{in},1}
+i g_c\cos{\theta} \sigma_{\tau}, \label{eq:aequivalent}\\
\dot{\sigma}_{\tau}  &=& -i \Delta_{\tau}\sigma_{\tau}-\gamma_c \sigma_{\tau}   +i g_c\cos{\theta} a\ \nonumber\\
&&+i \frac{B_x\cos{\theta}}{2}\sigma_{s}  ,\label{eq:tau-down}\\
\dot{\sigma}_{s}  &=&-i \Delta_{\mathrm{s}}\sigma_{s} +i  \frac{B_x\cos{\theta}}{2}\sigma_{\tau}
 , \label{eq:s-minus}
\end{eqnarray} 
where $\Delta_{\tau(\mathrm{s})}=\pm\frac{\Omega-B_z}{2}-E_0 -\omega_{\mathrm{R}}$.
As evident from these equations, although the  electric field of the cavity only couples to the charge excitation, the spin-charge hybridization generates an effective spin-photon coupling. 

\begin{figure}
\includegraphics[width=\columnwidth]{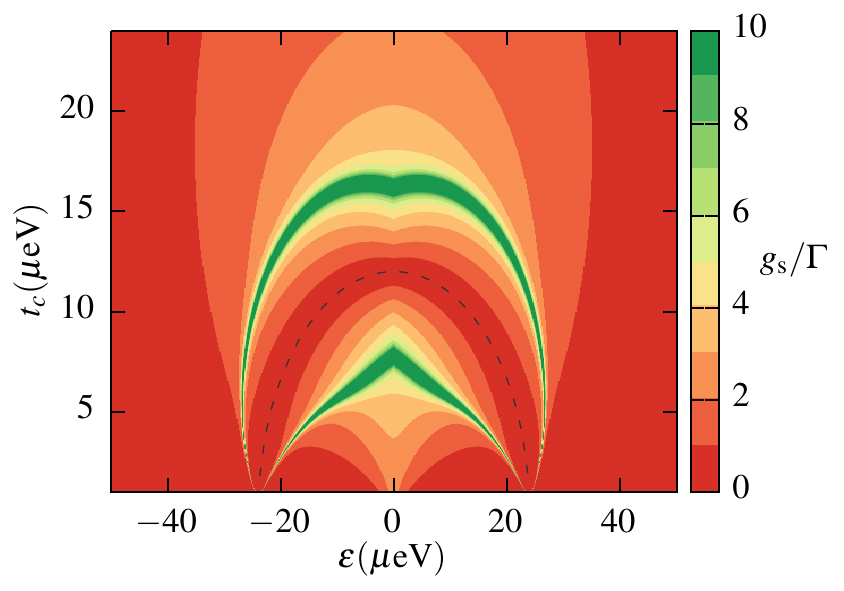}

\protect\caption{\label{fig:figure7}Strength of the spin-cavity coupling $g_{\mathrm{s}}/\Gamma$ according to Eq.~\eqref{eq:SC-condition} as a function of  $t_c$ and  $\epsilon$. $B_z$ has been adjusted to the resonance condition Eq.~\eqref{eq:resonance-condition}. 
Note that the strong-coupling regime ($g_{\mathrm{s}}/\Gamma>1$) is achieved away from the black dashed line,  $\Omega=\omega_{\mathrm{c}}=24\,\mu\mathrm{eV}$, in agreement with Fig.~\ref{fig:figure3}~(b).
The other parameters are   $B_x=1.62\,\mu\mathrm{eV}$,  $\gamma_1/2\pi=200\,\mathrm{MHz}$, $\gamma_{\phi}/2\pi=150|\sin{\theta}|\,\mathrm{MHz}$, $g_c/2\pi=40\,\mathrm{MHz}$, and $\kappa/2\pi=1.77\,\mathrm{MHz}$. }
\end{figure}

Solving equation Eq.~\eqref{eq:tau-down} for the steady state we can obtain the bright (B) mode that mediates this coupling, 
\begin{equation}
\sigma_{B}=\sin\alpha\,\sigma_{s} 
+ \cos{\alpha}\, a\ ,
\end{equation}
where we have introduced the angle $\alpha=\arctan\frac{B_x}{2g_{\mathrm{c}}}$.
For the Eqs.~\eqref{eq:aequivalent} and~\eqref{eq:s-minus}, we obtain 
the reduced dynamics 
\begin{eqnarray}
\dot{a}  &=&i (\Delta_0+\Delta_{\tau}\eta\cos^2\alpha) a-\frac{\kappa'}{2}  a+\sqrt{\kappa_1} a_{\mathrm{in},1} \nonumber\\
&&+i \sin\alpha\cos\alpha\,\eta(\Delta_{\tau}+i\gamma_{\mathrm{c}}) \sigma_{s}, \label{eq:areduced}\\
\dot{\sigma}_{s}  &=&-i( \Delta_{\mathrm{s}}-\Delta_{\tau}\eta\sin^2\alpha)\sigma_{s}-\gamma_s\sigma_{s} \nonumber\\ 
&&+i \sin\alpha\cos\alpha\,\eta(\Delta_{\tau}+i\gamma_{\mathrm{c}}) a
 , \label{eq:s-minusreduced}
\end{eqnarray} 
with
\begin{equation}
\eta=\frac{B_x^2/4+g_{\mathrm{c}}^2}{\Delta_{\tau}^2+\gamma_{\mathrm{c}}^2}\cos^2\theta ,
\end{equation}
and  effective decay rates  
\begin{eqnarray}
\kappa'& =&\kappa+2\gamma_{\mathrm{c}}\eta\cos^2\alpha=\kappa+2\gamma_{\mathrm{c}}\frac{g^2_{\mathrm{c}}\cos^2\theta}{\Delta_{\tau}^2+\gamma_{\mathrm{c}}^2},\\
\gamma_{\mathrm{s}} &=&\gamma_{\mathrm{c}}\eta\sin^2\alpha=\frac{\gamma_{\mathrm{c}}}{4}\frac{B_x^2\cos^2\theta}{\Delta_{\tau}^2+\gamma_{\mathrm{c}}^2}. 
\end{eqnarray}
According to the derivation in Appendix~\ref{sec:SC},
the resonance condition reads
\begin{equation}
(\Delta_s+\Delta_0)^{\mathrm{res}}=-\Delta_{\tau}\eta\frac{2\gamma_{\mathrm{c}}\eta+\kappa \cos{2\alpha}}{\kappa+2\gamma_{\mathrm{c}}\eta \cos{2\alpha}} \ ,\label{eq:resonance-condition}
\end{equation}
and  strong coupling is achieved for
\begin{equation}
g_{\mathrm{s}}>\Gamma\equiv\frac{|\kappa+2\gamma_{\mathrm{c}}\eta \cos{2\alpha}|}{2\sqrt{2}\sqrt{\frac{4\gamma_{\mathrm{c}}^2\eta^2+4\gamma_{\mathrm{c}}\eta\kappa\cos{2\alpha}+\kappa^2}{4\gamma_{\mathrm{c}}^2\eta^2+4\gamma_{\mathrm{c}}\eta\kappa\cos^2{\alpha}+\kappa^2}}} , \label{eq:SC-condition}
\end{equation}
where we have defined
\begin{equation}
g_{\mathrm{s}}=|\Delta_{\tau}|\eta\sin\alpha\cos\alpha=|\Delta_{\tau}|\frac{B_x g_{\mathrm{c}}\cos^2{\theta}}{2(\Delta_{\tau}^2+\gamma_{\mathrm{c}}^2)}.
\end{equation}
In Fig.~\ref{fig:figure7} we provide  a map of the coupling strength via the quantity $g_{\mathrm{s}}/\Gamma$, with the magnetic field  adjusted to the resonance condition.
This map  indicates the optimum working points that create a strong spin-photon interaction that overcomes the decoherence.

\section{Conclusions}
\label{sec:conclusions}

In conclusion, we detail the conditions for achieving  strong coupling between a single electron spin and a microwave cavity photon, which eventually would allow long distance spin-spin coupling and long-range spin-qubit gates. 
Non-local quantum gates may also facilitate quantum error correction within a fault-tolerant architecture, and have already been used for this purpose in other systems~\cite{Taylor2005,Reed2012,Waldherr2014}.
 
Our analysis on the dynamics of the full hybrid silicon-cQED system confirms that, 
with the recent advances in  Si DQDs fabrication and control, a spin-photon coupling of more than $10\,\mathrm{MHz}$ with a sufficiently low spin decoherence rate is achievable with this setup, potentially allowing the strong-coupling regime~\cite{Zajac2016,Mi2017b}.
In such a regime, the cavity not only can act as a mediator  of spin-spin coupling but also enables cavity-based readout of the spin qubit state~\cite{Kawakami2014,Takeda2016}.
Interestingly, the strong-coupling regime for the spin-cavity coupling may be attained even when the coupling strength of the charge-cavity coupling can not overcome the charge decoherence rate.   

Although here we have focused on the coupling of a single electron spin to a single photon, the implementation of proposals for other type of spin qubits with more than one electron \cite{Burkard2006,Jin2012,Taylor2013,Guido_RX_PRB_2015}  seems feasible with the present technology in Si QDs.

\begin{acknowledgments}
\textit{Acknowledgments.---}
This work has been supported by the Army Research Office grant W911NF-15-1-0149. 
Work at Princeton was also supported by 
 the U.S. Department of Defense under contract H98230-15-C0453, and the Gordon and Betty Moore Foundations EPiQS Initiative through grant GBMF4535.
\end{acknowledgments}

\appendix

\section{Multilevel RWA\label{sec:RWA}}

The time-dependent equations of motion, Eqs.~\eqref{eq:LLangevin1} and~\eqref{eq:LLangevin3}, can be solved within a rotating-wave-approximation (RWA) if the driving frequency is close to the transition energies of the system. Defining $\tilde{\sigma}_{n,n+j}=\sigma_{n,n+j}e^{i\omega_{\mathrm{R}} t}$ for $j>0$, these equations include both time-independent terms and  terms which oscillate at frequency $2\omega_{\mathrm{R}}$,
\begin{eqnarray}
\dot{a} & =&i \Delta_0 a-\frac{\kappa}{2} a+\sqrt{\kappa_1}a_{\mathrm{in},1}+\sqrt{\kappa_2}a_{\mathrm{in},2}\label{eq:LLangevin1-2}\\
&& -i g_c  \sum_{n=0}^{2}\sum_{j=1}^{3-n}d_{n,n+j}\tilde{\sigma}_{n,n+j}\nonumber\\
&& -i g_c  \sum_{n=1}^{3}\sum_{j=1}^{n}d_{n,n-j}\tilde{\sigma}_{n,n-j}e^{2i\omega_{\mathrm{R}} t} ,\nonumber\\
\dot{\tilde{\sigma}}_{n,n+j} & =&-i \left(E_{n+j}-E_n-\omega_{\mathrm{R}}\right) \tilde{\sigma}_{n,n+j}\label{eq:LLangevin3-2}\\
&&-\sum_{n'j'}\gamma_{n,n+j,n',n'+j'} \tilde{\sigma}_{n',n'+j'} \nonumber\\
&&-\sum_{n'j'}\gamma_{n,n+j,n',n'-j'} e^{2i\omega_{\mathrm{R}} t}\tilde{\sigma}_{n',n'-j'} \nonumber \\
&&+\sqrt{2\gamma}\mathcal{F}e^{i\omega_{\mathrm{R}} t}  -i g_c \left(a +a^{\dagger} e^{2i\omega_{R}t} \right)d_{n+j,n}\delta_{n,0}. \nonumber
\end{eqnarray} 
Here, $j,j'>0$. The RWA consists in neglecting the fast-oscillating, i.e., counter-rotating terms. For the mean value of the operators and using $\sigma_{n,n+j}$ instead of $\tilde{\sigma}_{n,n+j}$  to simplify the notation, the equations in the RWA read
 \begin{eqnarray}
\dot{a} & =&i \Delta_0 a-\frac{\kappa}{2} a+\sqrt{\kappa_1}a_{\mathrm{in},1}\label{eq:LLangevin1-3}\\
&& -i g_c  \sum_{n=0}^{2}\sum_{j=1}^{3-n}d_{n,n+j}\sigma_{n,n+j} , \nonumber\\
\dot{\sigma}_{n,n+j} & =&-i \left(E_{n+j}-E_n-\omega_{\mathrm{R}}\right) \sigma_{n,n+j}\nonumber\\
&&-\sum_{n'j'}\gamma_{n,n+j,n',n'+j'} \sigma_{n',n'+j'}  \label{eq:LLangevin3-3}\\
&&  -i g_{\mathrm{c}} a d_{n+j,n}\delta_{n,0}, \nonumber
\end{eqnarray} 
since $\left\langle a_{\mathrm{in},2}\right\rangle=0$ and $\left\langle\mathcal{F}\right\rangle=0$. In the equations, as in the main text, we have omitted the brackets  for simplicity.

\section{Decoherence model\label{sec:decoherence}}

We assume that the charge relaxation processes dominate over
direct spin relaxation.  
For the coupling to the phonon environment, the  Liouvillian superoperator   contains charge relaxation ($\gamma_{1}$) and pure dephasing ($\gamma_{\phi}$),
\begin{eqnarray}
\mathcal{L}_{\text{ph}}\rho&=&\frac{\gamma_1}{2}\left(2\sigma_-\rho\sigma_+-\sigma_+\sigma_-\rho-\rho\sigma_+\sigma_-\right) \nonumber\\
&+&\frac{\gamma_{\phi}}{4}\left(2\sigma_z\rho\sigma_z-\sigma_z\sigma_z\rho-\rho\sigma_z\sigma_z\right)
, \label{eq:decoherence}
\end{eqnarray}
where $\sigma_{\pm}=\left|\pm\right\rangle\left\langle \mp\right|$, $\sigma_z=\left|+\right\rangle\left\langle+\right|-\left|-\right\rangle\left\langle-\right|$. (Note that in this appendix the Pauli operators $\sigma_{\alpha}$ are in the basis of bonding and antibonding states, $\left|\pm\right\rangle$, instead of left and right.)

The interaction Hamiltonian for charge
decays can be written as 
\begin{equation}
H_{\mathrm{ph}} =  \sum_{\mathbf{k}} c_{\mathbf{k}} \left(|+\rangle\langle -| b_{\mathbf{k}} +|-\rangle\langle +|b_{\mathbf{k}}^\dagger \right),
\label{Hc}
\end{equation}
where $b_k$ annihilates a phonon in mode ${\mathbf{k}}$.
Therefore the relaxation rate at zero temperature can be obtained 
using Fermi's Golden Rule,
\begin{equation}
\gamma_{1} = \frac{2\pi}{\hbar}  \sum_f |\langle f| \langle -| H_{\mathrm{ph}}|+\rangle |0\rangle|^2 \delta(\Omega-E_f),
\label{relax1}
\end{equation}
where $|0\rangle$ and $|f\rangle$ are the initial phonon vacuum and single-phonon final states,  $\Omega$ is the orbital energy ($\Omega=\sqrt{\epsilon^2+4 t_c^2}$), and $E_f$ denotes the phonon energy.
Substituting Eq.~(\ref{Hc}) into Eq.~(\ref{relax1}), we obtain
\begin{equation}
\gamma_{1}  = \frac{2\pi}{\hbar}  \sum_{\mathbf{k}} |c_{\mathbf{k}}|^2 \delta(\Omega-E_{k})
 = \frac{2\pi}{\hbar}  |c_{k}|^2 D(\Omega),
\end{equation}
where $k$ is the modulus of the ${\mathbf{k}}$ vector evaluated at the energy $\Omega$.
Here, $D(E)$ is the phonon density of states.
In general, $\gamma_1$ depends on the parameters $t_c$ and $\epsilon$ both via $D(\Omega)$ and $c_k$, since $k=k(\Omega)$. We assume here a constant $\gamma_1$ since we expect that this approximation will hold in a small transition
energy window around the cavity frequency.

The pure dephasing term is due to charge noise or fluctuations in the value of $\epsilon$. The main contribution to the dephasing rate $\gamma_{\phi}$ is proportional to the first derivative of the energy transition with respect to $\epsilon$, i.e.,
\begin{equation}
\gamma_{\phi}\propto \frac{\partial (E_+-E_-)}{\partial \epsilon}=\sin{\theta},
\end{equation}
therefore is zero at the ``sweet spot'' $\epsilon=0$.

Using the  Liouvillian in Eq.~\eqref{eq:decoherence}, we can calculate the decoherence dynamics for the mean value of any operator as $\left\langle\dot{A}\right\rangle=\mathrm{tr}\{A\mathcal{L}_{\mathrm{ph}}\rho\}$. (In the following we omit the brackets for simplicity.) The coherences decay as 
\begin{equation}
\dot{\sigma}_{\pm}=- \gamma_{\mathrm{c}}\sigma_{\pm}=-\left( \frac{\gamma_1}{2} + \gamma_{\phi}\right)\sigma_{\pm}.\label{eq:tau}
\end{equation}

In this work we include the spin degree of freedom and here, we assume spin-independent rates. 
In the main text we have defined  
$\sigma_{\tau}=\left|-,\downarrow\rangle\langle+,\downarrow\right|$ and $\sigma_{s}=\left|-,\downarrow\rangle\langle -,\uparrow\right|$.  While $\sigma_{\tau}$  decays as
 $\dot{\sigma}_{\tau}=-\gamma_c \sigma_{\tau}$,
the same type of calculation taking into account the spin reveals 
$\dot{\sigma}_s=0$.
The decoherence part of the dynamics entering in Eq.~\eqref{eq:LLangevin3} is obtained via a rotation of the previous uncoupled equations into the eigenbasis of $H_{0}$, which results in
\begin{equation}
\dot{\sigma}_{nm}=-\sum_{n',m'}\gamma _{nm,n'm'} \sigma_{n'm'}.
\end{equation}
From the form of the eigenstates in the bonding-antibonding basis (Eqs.~\eqref{eq:vectors1} to~\eqref{eq:vectors4}),  we can determine the effect of charge dephasing in the eigenbasis of $H_{0}$. 
Since
\begin{eqnarray}
\sigma_{01} & \simeq \cos{\frac{\Phi}{2}}\sigma_s+\sin{\frac{\Phi}{2}}\sigma_{\tau}, \label{eq:01abbasis}\\
\sigma_{02} & \simeq \sin{\frac{\Phi}{2}}\sigma_s-\cos{\frac{\Phi}{2}}\sigma_{\tau}, \label{eq:02abbasis}
\end{eqnarray}
the decoherence dynamics  can be expressed as
\begin{eqnarray}
\begin{pmatrix}
\dot{\sigma}_{01} \\
\dot{\sigma}_{02} 
\end{pmatrix}\simeq-\gamma_c\begin{pmatrix}
\sin^2\frac{\Phi}{2} &  -\frac{\sin{\Phi}}{2} \\
-\frac{\sin{\Phi}}{2} &\cos^2\frac{\Phi}{2} 
\end{pmatrix}
\begin{pmatrix}
\sigma_{01} \\
\sigma_{02} 
\end{pmatrix}
.\label{eq:dmatrix}
\end{eqnarray}
Note also that $\sigma_{03}\simeq \left|-,\downarrow\rangle\langle+,\uparrow\right|$, therefore its decoherence is decoupled from the rest, $\dot{\sigma}_{03}\simeq-(\gamma_1+\gamma_{\phi})\sigma_{03}/2$.

\section{Characterization of the spin-photon coupling\label{sec:SC}}
 
 In this appendix, we detail the calculation of the strong-coupling condition for our two-level equivalent system. To examine this, we describe the spin-cavity interaction (Eqs.~\eqref{eq:areduced} and~\eqref{eq:s-minusreduced}) with an effective Hamiltonian in the one-excitation manifold. This means  that the Hilbert space is composed of states with a single excitation from the ground state, i.e., $\left\{\left|\downarrow,n+1\right\rangle, \left|\uparrow,n\right\rangle\right\}$, where $n$ is the number of photons in the cavity. In this basis, the effective Hamiltonian reads
 \begin{eqnarray}H_{\mathrm{eff}}&=&-
\begin{pmatrix}
\Delta_0+i\kappa'/2 &  0 \\
0 & -\Delta_{\mathrm{s}}+i\gamma_{\mathrm{s}}
\end{pmatrix}\\
&&-\eta\begin{pmatrix}
\Delta_{\tau}\cos^2\alpha &  (\Delta_{\tau}+i\gamma_{\mathrm{c}})\sin\alpha\cos\alpha \\
(\Delta_{\tau}+i\gamma_{\mathrm{c}})\sin\alpha\cos\alpha & \Delta_{\tau}\sin^2\alpha 
\end{pmatrix}
.\nonumber
\end{eqnarray} 
We note that the eigenvalues of $H_{\mathrm{eff}}$ are of the form
\begin{equation}
\mu_{\pm} = A - i B/2 \pm \sqrt{C + i D} ,
\end{equation}
with $A, B, C, D$ all real. $A$ can be seen to be the average frequency of the system, while $B$ is the average damping. However, the $C$ and $D$ terms are more subtle. For example, if one does not choose the  detunings such that $D \neq 0$, then the two vacuum Rabi-split peaks will have different linewidths. We also find a modification of the strong-coupling condition.

Let us now consider weak versus strong coupling for this system. For the usual Jaynes-Cummings model, where the two-level system is directly coupled to the cavity field, strong coupling is defined as sufficiently large interaction such that two separate peaks are observable in the photon response ($C > B^2/4$) \emph{and} such that the system modes are (near) equal combinations of matter and light ($C > 0, D = 0$).  For the Jaynes-Cummings case, the one-excitation Hamiltonian is 
 \begin{eqnarray}H_{\mathrm{JC}}&=&-
\begin{pmatrix}
\Delta_0+i\kappa/2 &  g_{JC} \\
g_{JC} & -\Delta_{JC}+i\gamma_{JC}
\end{pmatrix} ,
\end{eqnarray} 
and we have the combined condition 
\begin{equation}
g_{JC} > \sqrt{\frac{\gamma_{JC}^2 + (\kappa/2)^2}{2}} ,
\end{equation}
which reduces to the usual $g_{JC} > \gamma_{JC}, \kappa/2$ for $2\gamma_{JC} \approx \kappa$.

For our more unusual case, we will define resonance as $D = 0$ and strong coupling as $C > B^2/4$, as above.  We first examine the on-resonance condition. The detuning for resonance is 
\begin{equation}
(\Delta_s+\Delta_0)^{\mathrm{res}}=-\Delta_{\tau}\eta\frac{2\gamma_{\mathrm{c}}\eta+\kappa \cos{2\alpha}}{\kappa+2\gamma_{\mathrm{c}}\eta \cos{2\alpha}} \ .\label{eq:resonance-condition-app}
\end{equation}
This value corresponds to setting the detuning to match the (quantum) `Stark shifted' response of the spin and the photon.
We also find that strong coupling in the traditional sense arises for
\begin{equation}
g_{\mathrm{s}}>\Gamma\equiv\frac{|\kappa+2\gamma_{\mathrm{c}}\eta \cos{2\alpha}|}{2\sqrt{2}\sqrt{\frac{4\gamma_{\mathrm{c}}^2\eta^2+4\gamma_{\mathrm{c}}\eta\kappa\cos{2\alpha}+\kappa^2}{4\gamma_{\mathrm{c}}^2\eta^2+4\gamma_{\mathrm{c}}\eta\kappa\cos^2{\alpha}+\kappa^2}}} , \label{eq:SC-condition-app}
\end{equation} 
with the definition
\begin{equation}
g_{\mathrm{s}}=|\Delta_{\tau}|\eta\sin\alpha\cos\alpha=|\Delta_{\tau}|\frac{B_x g_{\mathrm{c}}\cos^2{\theta}}{2(\Delta_{\tau}^2+\gamma_{\mathrm{c}}^2)}.
\end{equation}

\bibliographystyle{apsrev4-1}
\bibliography{references}

\end{document}